# The dissociation of (a+c) misfit dislocations at the InGaN/GaN interface

J. Smalc-Koziorowska[1], J, Moneta[1], G. Muzioł[1], W. Chromiński[2], R. Kernke[3], M. Albrecht[3], T. Schulz[3] and I. Belabbas[4]

[1]Institute of High Pressure Physics, Polish Academy of Sciences, Warsaw 01142, Poland

[2]Faculty of Materials Science and Engineering, Warsaw University of Technology, Warsaw 02507, Poland

[3]Lebniz Institute for Crystal Growth, Berlin 12489, Germany

[4]Equipe de Cristallographie et de Simulation des Matériaux. Laboratoire de Physico-Chimie des Matériaux et Catalyse. Faculté des Sciences Exactes, Université de Bejaia 06000, Algeria.

## Abstract

**(a+c)** dislocations in hexagonal materials are typically observed to be dissociated into partial dislocations. Edge **(a+c)** dislocations are introduced into (0001) nitride semiconductor layers by the process of plastic relaxation. As there is an increasing interest in obtaining relaxed InGaN buffer layers for the deposition of high In content structures, the study of the dissociation mechanism of misfit **(a+c)** dislocations laying at the InGaN/GaN interface is then crucial for understanding their nucleation and glide mechanisms. In the case of the presented plastically relaxed InGaN layers deposited on GaN substrates we observe a trigonal network of **(a+c)** dislocations extending at the interface with a rotation of 3° from <1$\bar{1}$00> directions. High resolution microscopy studies show that these dislocations are dissociated into two Frank-Shockley 1/6<22,$\bar{0}$3> partial dislocations with the $I_1$ BSF spreading between them. Atomistic simulations of a dissociated edge **(a+c)** dislocation revealed a 3/5 atom ring structure for the cores of both partial dislocations. The observed separation between two partial dislocations must result from the climb of at least one of the dislocations during the dissociation process, possibly induced by the mismatch stress in the InGaN layer.

**Introduction**

(**a+c**) dislocations have the largest Burgers vector among perfect dislocations in a hexagonal structure. It has been theoretically predicted and experimentally proven that a dislocation with such a large Burgers vector tends to dissociate into partial dislocations to minimize its elastic energy. [1,2] The movement and dissociation mechanisms of 1/3 <11$\bar{2}$3> (**a+c**) dislocations are crucial for the mechanical properties of metals with a hexagonal structure like magnesium or titanium. Many theoretical and experimental works have treated the subject of the dissociation mechanism and glide of (**a+c**) dislocations in materials with hexagonal structures. [2-6] In the case of nitride semiconductors crystallizing in the wurtzite structure, the (**a+c**) dislocations are frequently found in layers deposited on highly mismatched substrates, such as GaN or AlN on sapphire or silicon,[7,8] and their origin is probably due to coalescence of slightly misoriented islands at early stages of an epilayer growth.[9] They also play an important role in the strain relaxation of nitride semiconductor layers crystallizing in the wurtzite structure and oriented along [0001] direction. It has been shown that $In_xGa_{1-x}N$ layers (InGaN) deposited on bulk GaN or GaN/sapphire templates relax by the introduction of (**a+c**) misfit dislocation half-loops arranged in a trigonal net in cases when other easy glide systems cannot be activated. [10,11] Such dislocations were also observed at the interface between relaxed $Al_xGa_{1-x}N$ deposited on GaN.[12]

Most of the available data in the literature concerning (**a+c**) dislocations in nitride semiconductor epitaxial layers is related to threading dislocations with mixed character since their line direction is roughly oriented along the [0001] growth directions of these layers.[7, 13-15] These reports show that typically 50% of the observed (**a+c**) dislocations are dissociated.[7, 13] Rhode et al showed that doping of the GaN layer may suppress the dissociation of (**a+c**) dislocations.[16] Atomistic simulations of the core of (**a+c**) mixed dislocations show that the stable undissociated core should keep 5/7- atom or 5/6 atom rings structures[17, 18] while the dissociated cores comprise two partial dislocations with 9-atom and 7-atom rings, or with 5-atom and 7-atom rings, separated by a Drum-like prismatic stacking fault.[7] There is much less information about edge (**a+c**) dislocations in nitride semiconductors in the literature. Chen et al.[19] analyzed core structures of (**a+c**) edge dislocations on the $\{11\bar{2}2\}$ plane in gallium nitride by employing molecular dynamics simulations and observed that the most favorable configuration for these dislocations has a dissociated core. Experimentally, (**a+c**) edge dislocations were only observed in relaxed nitride layers deposited along the [0001] wurtzite direction. They form by the introduction of a misfit dislocation half-loop at the surface and then glide to the interface on $\{11\bar{2}2\}$ planes.[10, 11] High resolution transmission electron microscopy images of such dislocations laying at the $In_{0.2}Ga_{0.8}N$/GaN interface clearly show the dissociation of the (**a+c**) dislocations as both partial dislocations are either separated along **a** or **c** direction.[20] As there is an increasing interest in obtaining relaxed InGaN buffer layers for the deposition of high In content structures, the study of the dissociation mechanism of misfit (**a+c**) dislocations laying at the InGaN/GaN interface is then crucial for understanding their nucleation and glide mechanisms.

**Materials & methods**

InGaN layers with indium content in the range of 18 - 20% and thicknesses in the range from 40 to 100 nm were investigated. The structures were deposited either by metalorganic chemical vapor deposition (MOVPE) or by molecular beam epitaxy (MBE). The studied structures were grown on substrates with various threading dislocation densities (TDD) such as bulk GaN substrates prepared by halide vapor phase epitaxy – HVPE- (TDD ~$10^6$ cm$^{-2}$) or by ammonothermal growth (TDD ~$10^4$ cm$^{-2}$). The miscut of the substrates was in the range of 0.3-0.5° towards $[11\bar{0}0]$ direction. We carried out structural studies employing transmission electron microscopy (TEM) and scanning-transmission electron microscopy (STEM) using an FEI Tecnai G2 F20 S-TWIN operated at 200 kV and a Thermo Fischer Scientific Spectra200, both equipped with a high-angle annular dark field (HAADF) STEM detectors. Observations were performed in cross-section along the <11$\bar{2}$0> and <1$\bar{1}$00> zone axes and in plan-view along [0001] direction. Specimen preparation was performed by mechanical polishing, followed by ion milling in the PIPS system from Gatan.

Cathodoluminescence (CL) experiments were carried out in the Thermo Fisher Apreo SEM equipped with a Gatan Monarc system. Image acquisition was done recording the monochromatic light (460 nm) emitted from the InGaN layer using an acceleration voltage of 3 keV and a probe current of 1.6 nA.

An atomistic model of the dissociated (**a+c**) edge dislocation was constructed as a supercell-cluster hybrid.[21, 22] This consists of a large rectangular parallelepiped cell with sides along [0001], [1$\bar{2}$10] and [10$\bar{1}$0] directions. The model contains about 48000 atoms and has the following dimensions $[60c \times 100a \times 1a\sqrt{3}]$, where **a** and **c** represent the lattice parameters of wurtzite GaN. The model is periodic along the dislocation line direction and finite in the transversal directions. Thus, periodic boundary conditions were applied along the [10$\bar{1}$0] direction where fixed boundary conditions were adopted perpendicularly to the dislocations line direction. Initial atomic positions were generated by applying the displacement field of an edge dislocation given by isotropic elasticity theory.[23] Equilibrium atomic positions were obtained through a relaxation procedure based on quenched molecular dynamics.[23, 24] The equilibrium is attained when the average thermodynamic temperature of the system becomes smaller than $10^{-6}$ K. Atoms inside a 145Å radius cylinder, around the dislocation line, were allowed to relax while the other ones were fixed to their initial positions. The atomic energies and forces were evaluated using a modified Stillinger-Weber potential,[25] which in addition to regular Ga-N bonds, additionally accounts for Ga-Ga and N-N wrong bonds which may occur in the core region of a dislocation.

**Results**

InGaN (0001) layers relax by the introduction of (**a+c**) misfit dislocations half-loops at the layer surface. Fig. 1a shows a scheme of a glide of (**a+c**) misfit dislocations on {11$\bar{2}$2} planes. Each {11$\bar{2}$2} plane is in mirror relation to other {11$\bar{2}$2} plane, like shown here ($\bar{2}$112) and ($\bar{2}$11$\bar{2}$) planes. The glide of the 1/3[2$\bar{1}\bar{1}$3] dislocation on the ($\bar{2}$112) plane and the 1/3[2$\bar{1}\bar{1}\bar{3}$] dislocation on the ($\bar{2}$11$\bar{2}$) plane results in the introduction of the parallel misfit segments propagating along the [01$\bar{1}$0] direction. Plan-view cathodoluminescence studies of a 50 nm thick $In_{0.2}Ga_{0.8}N$ layer deposited by MBE on bulk GaN substrate prepared by the HVPE method show a trigonal net of (**a+c**) misfit dislocations visible as dark lines (Fig. 1b). It can be noticed that each <1$\bar{1}$00> set of dislocations contains dislocations propagating at some angle to the <1$\bar{1}$00> direction. Careful studies of the dislocation configuration employing plan-view TEM investigations reveal that each set contains two subsets of misfit dislocations propagating at around ±3º from <1$\bar{1}$00> direction (Fig. 1c). This rotation angle is the same for all three sets of dislocations and neither dislocation set is true along <1$\bar{1}$00> directions. Similar rotation was observed for dislocations formed in cubic layers like SiGe [26] or InGaAs [27] deposited on miscut substrates.

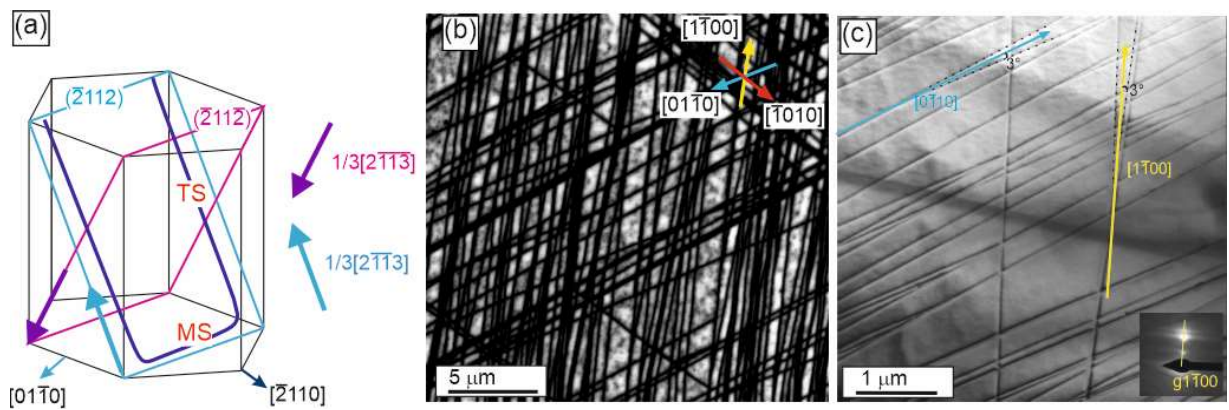

*Fig. 1 (a) Scheme of the glide of **(a+c)** dislocations on {11$\bar{2}$2} pyramidal planes. The dislocation loop consists of the threading screw segment (TS) and misfit edge segment (MS). (b) Cathodoluminescence map taken at wavelength 460 nm (3kV acceleration voltage, 1.6nA probe current),(c) Bright-field plan view TEM image taken with g1$\bar{1}$00 of the **(a+c)** dislocation network at the In$_{0.2}$Ga$_{0.8}$N/GaN interface, the 3 degrees rotation of dislocation lines from <1$\bar{1}$00> directions is indicated.*

In material systems where mirror-related glide planes are active, misfit dislocations introduced in these planes are parallel to each other (Fig. 1a). However, the substrate miscut changes the intersection line of the glide plane and interface and consequently the propagation direction of misfit dislocations. This phenomenon results in two sets of dislocations rotated by some angle depending on the miscut angle. However, simple geometrical calculations reveal that the {11$\bar{2}$2} slip system requires about 5° substrate miscut in order to result in a 6° rotation between misfit dislocations,[28] which exceeds by far the miscut used in this study in the range from 0.3-0.5° and no dependence of the rotation angle on the value of the substrate miscut angle is observed. Additionally, the rotation between misfit dislocations is the same for all three misfit dislocation sets, while it should affect mostly the projection of the misfit dislocations laying parallel to the miscut direction. Thus, the phenomenon of misfit dislocation rotation cannot be explained by a substrate miscut like in the case of cubic layers deposited on highly misoriented substrates. Alternatively, the inclination of the dislocation line in the glide plane can also be excluded since cross-section studies show that all misfit dislocations lay at the InGaN/GaN interface on the (0001) plane (Fig.2a). The elucidation of this unusual rotation of the misfit dislocation lines requires then further studies.

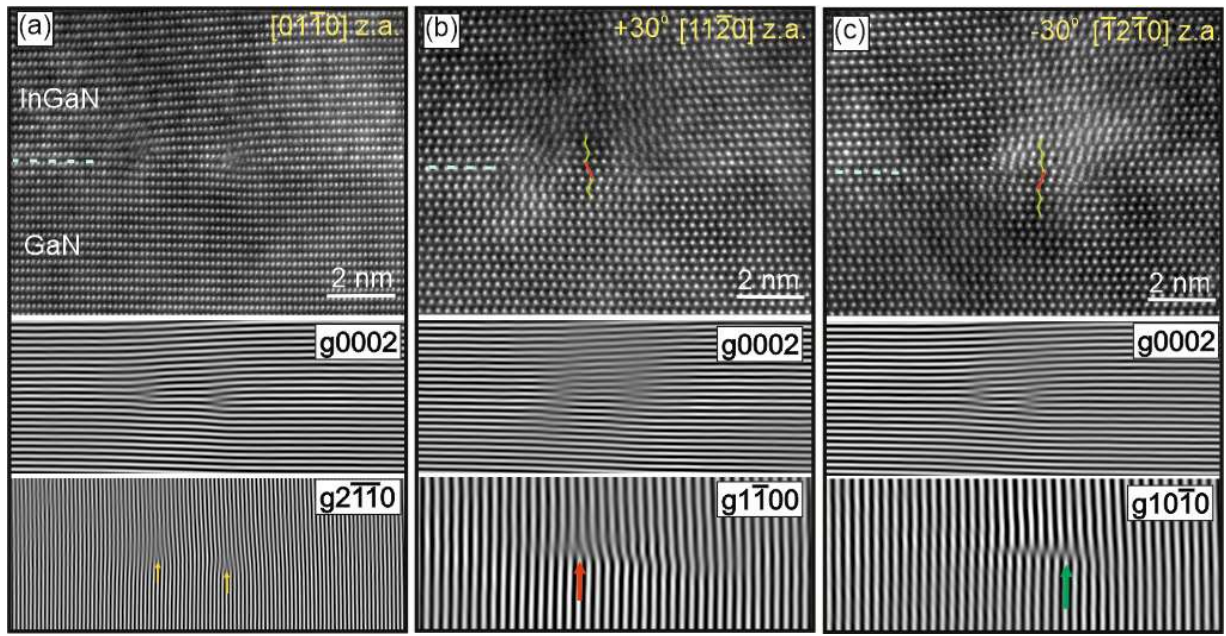

*Figure 2. HRSTEM images of the dissociated core of the misfit edge segment of an (**a+c**) dislocation at the $In_{0.2}Ga_{0.8}N$/GaN interface. (a) edge-on image of the dislocation core taken along $[01\bar{1}0]$ zone axis, (b) image of the (**a+c**) dislocation taken after the rotation of 30 degrees to $[\bar{1}2\bar{1}0]$ zone axis, and (c) after the rotation of 30 degrees to $[11\bar{2}0]$ zone axis. The stacking sequence showing the position of the $I_1$ basal stacking fault is indicated with red lines. The corresponding Bragg images made using frequencies indicated in the images show the position of the extra half-planes corresponding to the position of partial dislocations.*

HRSTEM studies of the misfit segments of (**a+c**) dislocations show the dissociated core of such dislocations spreading over several nanometers along the **a** direction and also separated along the **c** direction (Fig. 2a). The contrast of the atomic columns inside the cores of the partial dislocations is a little bit smeared, which may be due to observed 3 degrees rotation of dislocation lines. The Bragg filtering of Fig. 2a made using in-plane $2\bar{1}\bar{1}0$ and out-of-plane 0002 frequencies of the Fast Fourier Transform (FFT) clearly show the presence of two extra ($2\bar{1}\bar{1}0$) planes and correspondingly two extra (0002) planes, indicating a dissociation of the 1/3 $[2\bar{1}\bar{1}3]$ dislocation according to the reaction:

$$1/3\,[2\bar{1}\bar{1}3] \rightarrow 1/6\,[2\bar{1}\bar{1}3] + 1/6\,[2\bar{1}\bar{1}3] \tag{1}$$

No intersection between the (0002) and the ($2\bar{1}\bar{1}0$) additional half-planes from different partial dislocations is observed in dissociated (**a+c**) edge dislocations. The separation distance along [$2\bar{1}\bar{1}0$] direction for the presented here dislocation is 17 x ($2\bar{1}\bar{1}0$) planes which is ~27 Å, while the average distance measured for 15 different dissociated (**a+c**) dislocations is 24.17 Å. The separation between the partial dislocations along the **a** direction indicates the formation of a stacking fault placed between the dislocations on the basal plane. However, the ABAB stacking of the wurtzite lattice cannot be evidenced in the imaging along <$1\bar{1}00$> zone axes. The HRSTEM investigation along two neighboring <$11\bar{2}0$> zone axes shows a presence of the $I_1$ basal stacking fault (BSF) spreading between the partial dislocations. In both <$11\bar{2}0$> zone axes only one extra half-plane is visible in Bragg images made using in-plane frequencies of FFT, which is ($1\bar{1}00$) half-plane in Fig. 2b and ($10\bar{1}0$) half-plane in Fig.2c, indicating the presence of partial dislocations with Burgers vectors accordingly 1/6[$2\bar{2}03$] (Fig.2b) and 1/6[$20\bar{2}3$] (Fig. 2c). The formation of $I_1$ BSF requires then an additional shift of each 1/6[$2\bar{1}\bar{1}3$] partial dislocations along [$01\bar{1}0$] direction according to the reactions:

$$1/6\,[2\bar{1}\bar{1}3] + 1/6\,[01\bar{1}0] \rightarrow 1/6\,[20\bar{2}3] \tag{2}$$

$$1/6\,[2\bar{1}\bar{1}3] - 1/6\,[01\bar{1}0] \rightarrow 1/6\,[2\bar{2}03] \tag{3}$$

Then the dissociation reaction of this (**a+c**) dislocation may be described as follows:

$$1/3\,[2\bar{1}\bar{1}3] \rightarrow 1/6\,[20\bar{2}3] + 1/6\,[2\bar{2}03] + I_1\,BSF \tag{4}$$

Where the perfect **(a+c)** dislocation dissociates into two Frank-Shockley 1/6 <20$\bar{2}$3> dislocations terminating the $I_1$ BSF.

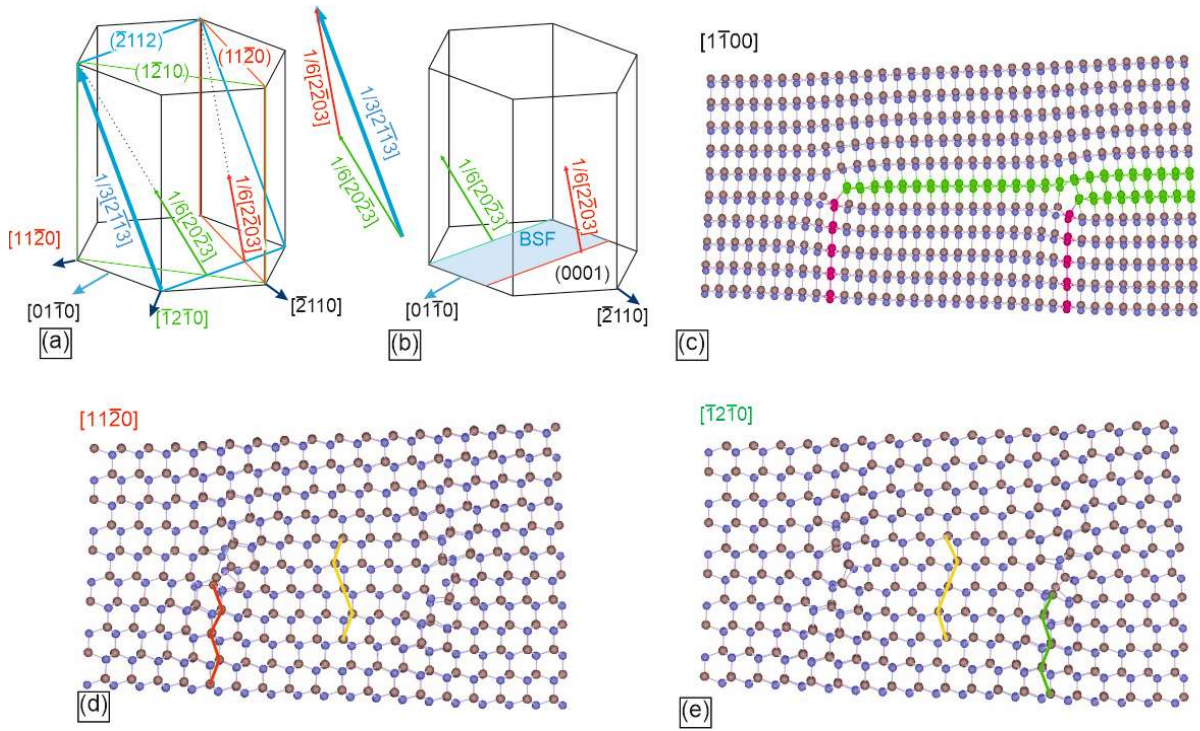

*Figure 3. (a) Scheme showing the relation between the perfect **(a+c)** dislocation (blue) and Frank-Shockley partial dislocations (green and red) on the ($\bar{2}$112) plane, (b) scheme of the dissociated **(a+c)** dislocation with the $I_1$ BSF spreading between the two partial dislocations on the (0001) plane; (c) Atomistic simulation of the dissociated core structure of the **(a+c)** dislocation presented in Fig. 2a viewed along [01$\bar{1}$0] zone axis, two extra **c** half-planes and **a** half-planes are indicated with green and red color respectively; (d) view of the simulated structure after the rotation to [11$\bar{2}$0] zone axis, (e) view of the simulated structure after the rotation to [$\bar{1}$2$\bar{1}$0] zone axis. The position of the Frank-Shockley partial dislocation is indicated with the red and the green line, respectively. The stacking of the $I_1$ BSF is indicated with the yellow line.*

Since the observed 3° rotation of (**a+c**) dislocation lines makes impossible the exact visualization of the core structure of partial dislocations, we performed atomistic simulations of the dissociated structure of the (**a+c**) dislocation. The core structure of the dissociated (**a+c**) edge dislocation at equilibrium, obtained via large-scale atomistic simulation, is represented in Fig.3c. It shows the (**a+c**) edge dislocation dissociated into two Frank-Shockley partial dislocations, which are 73.3° mixed dislocations, separated by a ribbon of an $I_1$ BSF. As pointed out previously, the $I_1$ BSF stacking cannot be observed along the <10$\bar{1}$0> direction but rather along <11$\bar{2}$0> direction. However, a rotation of the model by ± 30° around the [0001] axis makes the $I_1$ BSF appear between the two partials (Fig.3.d-e). The two Frank-Shockley partial dislocations have the same core structure. According to the terminology introduced in our previous report, [29] the dislocations have cores with a 3/5-atom ring structure. Depending on the nature of the atomic column at the termination of the additional (2$\bar{1}\bar{1}$0) half-plane, the core of the Frank-Shockley partial can be either with gallium or nitrogen polarity where, respectively, Ga-Ga or N-N wrong bonds are established.

As was already mentioned, there is a large separation between the partial dislocations along the in-plane direction (Fig. 2a), and each observed in this study (**a+c**) dislocation exhibits such a separation. The presented model shows no separation between extra (0002) half-planes along [0001] direction (Fig. 3c). However, we also observe such configurations of the dissociated core where the partial dislocations are separated along [0001] direction with a typically odd number of (0002) planes. It indicates that the dissociation mechanism of the (**a+c**) misfit dislocation at the GaN/InGaN interface must involve climb of the partial dislocations along <11$\bar{2}$0> and [0001] directions.

**Discussion**

The molecular dynamics simulations conducted by Chen et al. [19] of the core of the edge (**a+c**) dislocation on the {11$\bar{2}$2} pyramidal plane in GaN showed that the perfect (**a+c**) dislocation tends to dissociate into two partial dislocations. They identified a possible dissociation reaction as 1/3[11$\bar{2}$3] = (1-λ)/6[11$\bar{2}$3] + λ/6[11$\bar{2}$3] with the dissociation coefficient λ being around 0.5. A similar dissociation reaction was observed for the threading mixed dislocations observed in GaN deposited on sapphire.[14, 18] In their simulation, both partial dislocations stay in close vicinity on the {11$\bar{2}$2} plane, and they found that the average spacing between these dislocations is in the range of 3.4Å or 4.7Å on the glide plane, depending on the primary position of perfect dislocations on the {11$\bar{2}$2} plane. The structure predicted by Chen et al. is essentially different than presented here results, where both partial dislocations are separated along **a** and **c** directions and are connected by the I$_1$ BSF. However, the misfit dislocations laying at InGaN/GaN interface are under high misfit stress, and probably its dissociated structure is governed by the mismatch relaxation process and the minimization of dislocation's elastic strain energy.

The observed configuration of dissociated edge (**a+c**) dislocations resembles more the structure of dissociated dislocations found by molecular dynamics in plastically deformed magnesium.[2, 5] Molecular dynamics simulation carried out by Wu and Curtin[2] shows that the (**a+c**) dislocation dissociates first into two ½(**a+c**) partial dislocations on the pyramidal {11$\bar{2}$2} plane. Depending on the applied stress, the simulation led to three different configurations of the core including separation into two Frank-Shockley partial dislocations with I$_1$ BSF extended between them. Tang et al.[5] showed that dissociation of (**a+c**) dislocation into two Frank-Shockley partial dislocations is achieved by separate movements of atoms on the glide plane with different slip vectors of 1/6 [1$\bar{1}$00] and 1/6 [$\bar{1}$100]. Albeit wurtzite nitride semiconductors have the same symmetry as magnesium, the bonds in nitrides have ionic/covalent character so they cannot be directly compared to metals. Nevertheless, the results presented here suggest some common rules for the dissociation mechanism of the (**a+c**) dislocations in materials with hexagonal structures.

It should be also noted again that the presented here edge (**a+c**) dislocations are placed at the mismatched interface so mismatch stress also influences the dissociation process. Possibly the observed rotation of the misfit dislocation lines from <1$\bar{1}$00> direction may also be related to the dissociation mechanism. Nevertheless, a full understanding of the phenomena requires further studies.

**Conclusion**

Edge (**a+c**) dislocations can be introduced into (0001) nitride semiconductor layers by the process of plastic relaxation. (**a+c**) dislocation half-loops are formed at the layer surface and glide to the interface on {11$\bar{2}$2} planes. In the case of the presented InGaN layers deposited on GaN substrates we observe a trigonal network of (**a+c**) dislocations extending at the interface with a 3° rotation from <1$\bar{1}$00> directions. High resolution microscopy studies show that these dislocations are dissociated into two partial dislocations and separated along <11$\bar{2}$0> and [0001] directions. Imaging along <11$\bar{2}$0> zone axes reveals that both partial dislocations are Frank-Shockley 1/6<2$\bar{2}$03> partial dislocations with the I$_1$ BSF spreading between them. Atomistic simulations of the dissociated (**a+c**) dislocation demonstrated that the resulting Frank-Shockley partials have a core with a 3/5-atom ring structure. The observed separation between two partial dislocations must result from the climb of at least one of the dislocations during the dissociation process, possibly induced by the mismatch stress in the InGaN layer.

## Acknowledgments

This work was funded in part by Deutsche Forschungsgemeinschaft, Germany, project 465219948, and National Science Center, Poland, project OPUS LAP 2020/39/I/ST5/03379. For the purpose of Open Access, the author has applied a CC-BY public copyright license to any Author Accepted Manuscript (AAM) version arising from this submission.

## References

1. D.Hull & Bacon, D. J. (2001) *Introduction to dislocations,* Butterworth-Heinemann, Oxford.
2. Wu, Z. & Curtin, W. A. (2015) The origins of high hardening and low ductility in magnesium. *Nature,* **526,** 62-67.
3. Itakura, M., Kaburaki, H., Yamaguchi, M. & Tsuru, T. (2016) Novel Cross-Slip Mechanism of Pyramidal Screw Dislocations in Magnesium. *Physical Review Letters,* **116**.
4. Li, M., Tian, X., Jiang, W., Wang, Q. & Fan, H. (2023) Mechanism of strain hardening of magnesium single-crystals: Discrete dislocation dynamics simulations. *Journal of the Mechanics and Physics of Solids,* **173,** 105238.
5. Tang, X.-Z., Guo, Y.-F., Xu, S. & Wang, Y.-S. (2015) Atomistic study of pyramidal slips in pure magnesium single crystal under nano-compression. *Philosophical Magazine,* **95,** 2013-2025.
6. Clouet, E., Caillard, D., Chaari, N., Onimus, F. & Rodney, D. (2015) Dislocation locking versus easy glide in titanium and zirconium. *Nature Materials,* **14,** 931-936.
7. Rhode, S. L., Horton, M. K., Sahonta, S. L., Kappers, M. J., Haigh, S. J., Pennycook, T. J., McAleese, C., Humphreys, C. J., Dusane, R. O. & Moram, M. A. (2016) Dislocation core structures in (0001) InGaN. *Journal of Applied Physics,* **119**.
8. Sugawara, Y., Ishikawa, Y., Watanabe, A., Miyoshi, M. & Egawa, T. (2016) Analysis of reaction between c+a and -c+a dislocations in GaN layer grown on 4-inch Si(111) substrate with AlGaN/AlN strained layer superlattice by transmission electron microscopy. *AIP Advances,* **6,** 045020.
9. Kehagias, T., Komninou, P., Nouet, G., Ruterana, P. & Karakostas, T. (2001) Misfit relaxation of the AlN/Al2O3 (0001) interface. *Physical Review B,* **64,** 195329.
10. Moneta, J., Siekacz, M., Grzanka, E., Schulz, T., Markurt, T., Albrecht, M. & Smalc-Koziorowska, J. (2018) Peculiarities of plastic relaxation of (0001) InGaN epilayers and their consequences for pseudo-substrate application. *Applied Physics Letters,* **113,** 031904.
11. Srinivasan, S., Geng, L., Liu, R., Ponce, F. A., Narukawa, Y. & Tanaka, S. (2003) Slip systems and misfit dislocations in InGaN epilayers. *Applied Physics Letters,* **83,** 5187-5189.
12. Floro, J. A., Follstaedt, D. M., Provencio, P., Hearne, S. J. & Lee, S. R. (2004) Misfit dislocation formation in the AlGaN/GaN heterointerface. *Journal of Applied Physics,* **96,** 7087-7094.
13. Massabuau, F. C. P., Rhode, S. L., Horton, M. K., O'Hanlon, T. J., Kovács, A., Zielinski, M. S., Kappers, M. J., Dunin-Borkowski, R. E., Humphreys, C. J. & Oliver, R. A. (2017) Dislocations in AlGaN: Core Structure, Atom Segregation, and Optical Properties. *Nano Letters,* **17,** 4846-4852.
14. Hirsch, P. B., Lozano, J. G., Rhode, S., Horton, M. K., Moram, M. A., Zhang, S., Kappers, M. J., Humphreys, C. J., Yasuhara, A., Okunishi, E. & Nellist, P. D. (2013) The dissociation of the a plus c dislocation in GaN. *Philosophical Magazine,* **93,** 3925-3938.
15. Lozano, J. G., Yang, H., Guerrero-Lebrero, M. P., D'Alfonso, A. J., Yasuhara, A., Okunishi, E., Zhang, S., Humphreys, C. J., Allen, L. J., Galindo, P. L., Hirsch, P. B. & Nellist, P. D. (2014)

| | |
|---|---|
| | Direct Observation of Depth-Dependent Atomic Displacements Associated with Dislocations in Gallium Nitride. *Physical Review Letters,* **113,** 135503. |
| 16 | Rhode, S. K., Horton, M. K., Kappers, M. J., Zhang, S., Humphreys, C. J., Dusane, R. O., Sahonta, S. L. & Moram, M. A. (2013) Mg Doping Affects Dislocation Core Structures in GaN. *Physical Review Letters,* **111,** 025502. |
| 17 | Belabbas, I., Bere, A., Chen, J., Petit, S., Belkhir, M. A., Ruterana, P. & Nouet, G. (2007) Atomistic modeling of the (a plus c)-mixed dislocation core in wurtzite GaN. *Physical Review B,* **75,** 115201. |
| 18 | Horton, M. K., Rhode, S. L. & Moram, M. A. (2014) Structure and electronic properties of mixed (a + c) dislocation cores in GaN. *Journal of Applied Physics,* **116,** 063710. |
| 19 | Chen, C., Meng, F. & Song, J. (2015) Core structures analyses of (a+c)-edge dislocations in wurtzite GaN through atomistic simulations and Peierls–Nabarro model. *Journal of Applied Physics,* **117,** 194301. |
| 20 | Moneta, J., Grzanka, E., Turski, H., Skierbiszewski, C. & Smalc-Koziorowska, J. (2020) Stacking faults in plastically relaxed InGaN epilayers. *Semiconductor Science and Technology,* **35,** 034003. |
| 21 | Belabbas, I., Dimitrakopulos, G., Kioseoglou, J., Bere, A., Chen, J., Komninou, P., Ruterana, P. & Nouet, G. (2006) Energetics of the 30 degrees Shockley partial dislocation in wurtzite GaN. *Superlattices and Microstructures,* **40,** 458-463. |
| 22 | Belabbas, I., Chen, J., Akli Belkhir, M., Ruterana, P. & Nouet, G. (2006) New core configurations of the c -edge dislocation in wurtzite GaN. *physica status solidi c,* **3,** 1733-1737. |
| 23 | Hirth, J. P. & Lothe, J. (1982) *Theory of Dislocations,* Krieger Publishing Company. |
| 24 | Verlet, L. (1967) Computer "Experiments" on Classical Fluids. I. Thermodynamical Properties of Lennard-Jones Molecules. *Physical Review,* **159,** 98-103. |
| 25 | Béré, A. & Serra, A. (2002) Atomic structure of dislocation cores in GaN. *Physical Review B,* **65,** 205323. |
| 26 | Bolkhovityanov, Y. B., Deryabin, A. S., Gutakovskii, A. K. & Sokolov, L. V. (2010) Formation of edge misfit dislocations in GexSi1−x(x∼0.4–0.8) films grown on misoriented (001)→(111) Si substrates: Features before and after film annealing. *Journal of Applied Physics,* **107,** 123521. |
| 27 | Kightley, P., Goodhew, P. J., Bradley, R. R. & Augustus, P. D. (1991) A mechanism of misfit dislocation reaction for GaInAs strained layers grown onto off-axis GaAs substrates. *Journal of Crystal Growth,* **112,** 359-367. |
| 28 | Moneta, J. (2023) Mechanisms of strain relaxation in InGaN films grown on (0001)-oriented GaN substrates.PhD Thesis Institute of High Pressure Physics, Polish Academy of Sciences, Warsaw. |
| 29 | Vasileiadis, I. G., Belabbas, I., Bazioti, C., Smalc-Koziorowska, J., Komninou, P. & Dimitrakopulos, G. P. (2021) Stacking Fault Manifolds and Structural Configurations of Partial Dislocations in InGaN Epilayers. *physica status solidi (b),* **258,** 2100190. |